\documentclass[sigconf]{acmart}

\setcopyright{none}
\renewcommand\footnotetextcopyrightpermission[1]{}

\usepackage{algorithm}
\usepackage{algpseudocode}
\usepackage{multirow}
\usepackage{caption}

\AtBeginDocument{%
  }

\title{WPGRec: Wavelet Packet Guided Graph Enhanced Sequential Recommendation}

\author{Peilin Liu}
\affiliation{%
  \institution{College of Software, Jilin University}
  \city{Changchun}
  \state{Jilin}
  \country{China}}
\email{liupl9922@mails.jlu.edu.cn}

\author{Zhiquan Ji}
\affiliation{%
  \institution{College of Software, Jilin University}
  \city{Changchun}
  \state{Jilin}
  \country{China}}
\email{jizq1622@mails.jlu.edu.cn}

\author{Gang Yan}
\affiliation{%
  \institution{College of Computer Science and Technology, Jilin University}
  \city{Changchun}
  \state{Jilin}
  \country{China}}
\email{gyan8@jlu.edu.cn}
\authornote{Corresponding author.}

\renewcommand{\shortauthors}{Peilin Liu, Zhiquan Ji, and Gang Yan}

\begin{document}

\begin{abstract}
Sequential recommendation aims to model users’ evolving interests from noisy and non-stationary interaction streams, where long-term preferences, short-term intents, and localized behavioral fluctuations may coexist across temporal scales. Existing frequency-domain methods mainly rely on either global spectral operations or filter-based wavelet processing. However, global spectral operations tend to entangle local transients with long-range dependencies, while filter-based wavelet pipelines may suffer from temporal misalignment and boundary artifacts during multi-scale decomposition and reconstruction. Moreover, collaborative signals from the user--item interaction graph are often injected through scale-inconsistent auxiliary modules, limiting the benefit of jointly modeling temporal dynamics and structural dependencies. To address these issues, we propose \textbf{W}avelet \textbf{P}acket \textbf{G}uided Graph Enhanced Sequential \textbf{Rec}ommendation (\textbf{WPGRec}), a unified time--frequency and graph-enhanced framework that aligns multi-resolution temporal modeling with graph propagation at matching scales. WPGRec first applies a full-tree undecimated stationary wavelet packet transform to generate equal-length, shift-invariant subband sequences. It then performs subband-wise interaction-graph propagation to inject high-order collaborative information while preserving temporal alignment across resolutions. Finally, an energy- and spectral-flatness-aware gated fusion module adaptively aggregates informative subbands and suppresses noise-like components. Extensive experiments on four public benchmarks show that WPGRec consistently outperforms sequential and graph-based baselines, with particularly clear gains on sparse and behaviorally complex datasets, highlighting the effectiveness of band-consistent structure injection and adaptive subband fusion for sequential recommendation.
\end{abstract}

\keywords{Sequential Recommendation, Wavelet Packet Transform, Subband Modeling, Graph-based Learning}

\maketitle
\pagestyle{plain}
\section{Introduction}

Recommender systems are one of the core tasks in information retrieval, where the goal is to rank candidate items according to their relevance to a user's current needs~\cite{ricci2022chapter1,he2017ncf,lu2012recommender}. In many real-world platforms, this ranking process is typically driven by user interaction sequences, in which behavioral signals evolve continuously over time and often exhibit a mixture of stable preferences, recurring patterns, and short-term interest shifts. Meanwhile, these interactions also naturally induce a user--item bipartite graph that contains high-order collaborative structure beyond any single behavior sequence~\cite{wang2019kgat,he2020lightgcn}. Therefore, an effective sequential recommender should not only capture dynamic preferences at different temporal scales, but also incorporate high-order collaborative signals without disrupting the temporal structure of the sequence.

A large body of sequential recommendation methods focuses on modeling temporal dependencies in the time domain. Early models, such as GRU4Rec~\cite{hidasi2015session}, NARM~\cite{li2017neural}, and STAMP~\cite{liu2018stamp}, mainly rely on recurrent structures or attention mechanisms to capture user intent in short sessions. Subsequently, convolution-based models such as Caser~\cite{tang2018personalized} and NextItNet~\cite{yuan2019simple} further enhanced the modeling of local patterns and long-range dependencies. Later, Transformer-based methods, including SASRec~\cite{kang2018self}, TiSASRec~\cite{li2020time}, BERT4Rec~\cite{sun2019bert4rec}, and BSARec~\cite{shin2024attentive}, further improved contextual sequence modeling through self-attention and related inductive biases. Although these methods have achieved strong performance, they typically encode multiple temporal factors within a single feature stream. When observed behaviors are sparse, truncated, or corrupted by noise, long-term and low-frequency signals are more likely to dominate the final representation, thereby weakening the model's ability to preserve short-term variations and localized patterns, even though such information is often crucial for next-item ranking.

To alleviate the above limitations, recent studies have begun to introduce signal-processing perspectives into sequential recommendation, with the aim of better characterizing the non-stationary nature of user behavior. FEARec~\cite{du2023frequency} enhances sequence modeling in the frequency domain and explicitly captures periodic preference patterns, while WaveRec~\cite{HeoKim2025WaveRec} further shows that, compared with purely Fourier-based operations, wavelet transforms are more suitable for modeling localized multi-resolution dynamics. Despite the promise of this direction, most existing methods still remain largely sequence-side designs. In particular, structural signals from the interaction graph are usually introduced only after temporal encoding, or fused through auxiliary modules that lack explicit alignment with the underlying frequency decomposition. As a result, the model may still mix information from different scales during propagation and fusion, which not only weakens the benefits of multi-resolution modeling, but also makes boundary noise more difficult to control effectively.

\begin{figure}[t]
  \centering
  \includegraphics[width=\linewidth,trim=8 8 8 8,clip]{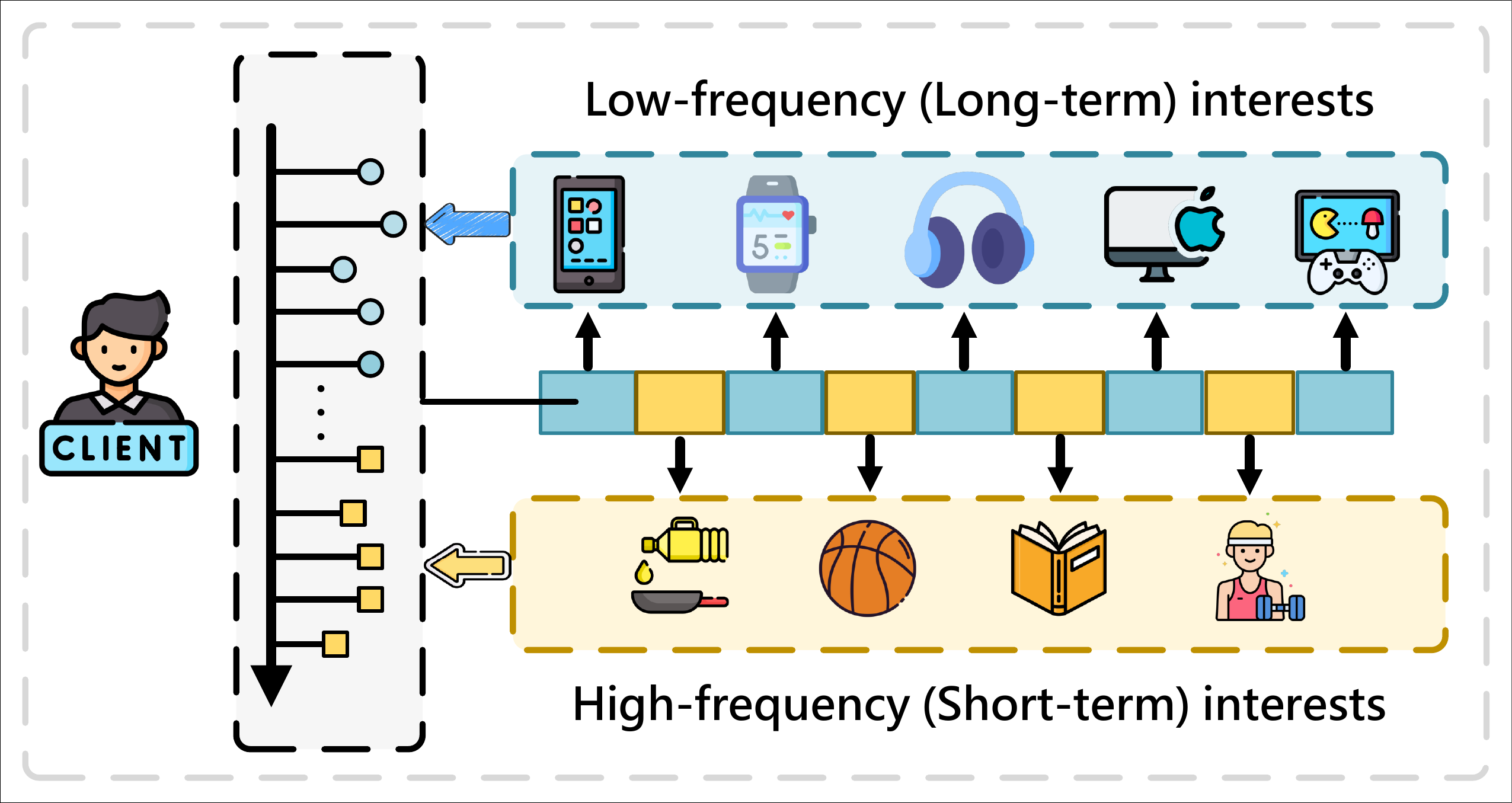}
  \caption{User behavior in sequential recommendation often involves both long-term preferences and short-term interests. This inherent multi-scale nature motivates us to explicitly disentangle and model preference signals at different temporal scales.}
  \label{fig:intro_motivation}
\end{figure}

In parallel, graph-based recommendation methods have demonstrated that collaborative structure is highly valuable for ranking tasks. Models such as PinSage~\cite{ying2018graph}, SR-GNN~\cite{wu2019session}, GC-SAN~\cite{xu2019graph}, SGL~\cite{wu2021self}, TGSRec~\cite{fan2021continuous}, and DGRec~\cite{yang2023dgrec} improve recommendation quality by propagating information over interaction graphs or graph-enhanced sequential structures. However, these methods usually do not explicitly distinguish sequential dynamics at different temporal scales. As illustrated in Figure~\ref{fig:intro_motivation}, user behavior in sequential recommendation often contains both long-term preferences and short-term interests. When long-term preferences, medium-range transition patterns, and short-term fluctuations are entangled within the same representation, graph propagation alone may still be insufficient. More importantly, if graph signals are fused with temporal features only at a coarse-grained level, the resulting representation may still suffer from scale mismatch.

These observations point to a simple but important principle: multi-resolution temporal modeling and graph propagation should be aligned at the same scale. Compared with downsampled wavelet pipelines, the Stationary Wavelet Packet Transform (SWPT) produces equal-length and shift-invariant subbands, making it particularly suitable for temporally aligned propagation and subband-level fusion. With this property, collaborative structure can be injected within each subband, rather than only after the temporal encoder has already mixed signals from different resolutions. Such a design is especially important for sequential recommendation, where ranking quality often depends on preserving both fine-grained local variations and stable long-range preferences.

Motivated by this idea, we propose \textbf{W}avelet \textbf{P}acket \textbf{G}uided Graph Enhanced Sequential \textbf{Rec}ommendation (\textbf{WPGRec}), a unified framework that combines scale-aligned time--frequency decomposition with band-consistent graph propagation. It applies a full-tree Stationary Wavelet Packet Transform (SWPT)~\cite{coifman1992entropy,nason1995swt,mallat2009wavelet} to decompose each behavior sequence into equal-length subbands that preserve temporal alignment across resolutions. On top of these aligned subbands, WPGRec further performs temporal aggregation and Chebyshev-polynomial-based graph propagation~\cite{defferrard2016cheby,KimEtAl2025ChebyCF}, so that collaborative information is injected within each subband rather than after different scales have already been mixed. Finally, we design an energy- and spectral-flatness-aware gating mechanism to adaptively fuse subband representations, emphasizing more informative components while suppressing spectra that are closer to noise~\cite{herre2001robust}. In this way, WPGRec reduces cross-scale interference, preserves localized temporal structure, and achieves more stable ranking performance across datasets with different sparsity levels and sequence characteristics.


\begin{figure*}[t]
  \centering
  \includegraphics[width=0.95\textwidth]{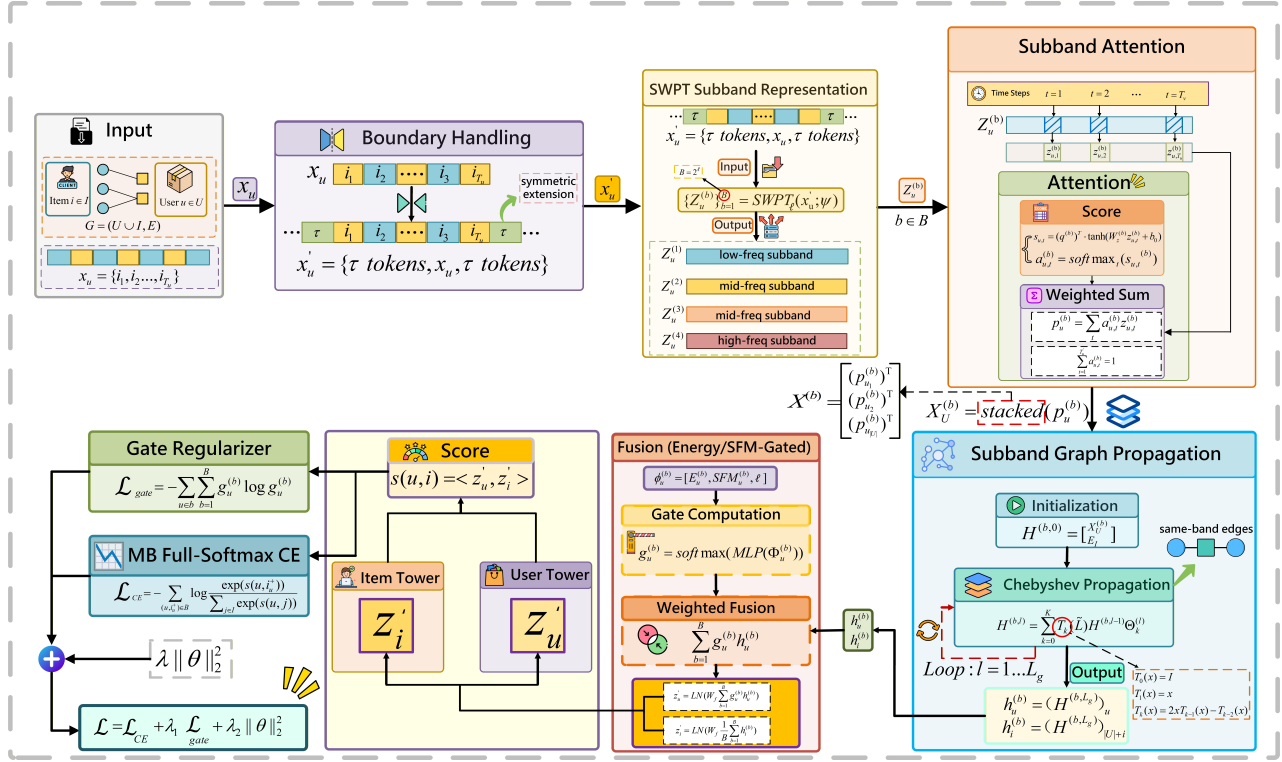}
  \caption{Overall Architecture of WPGRec.}
  \label{fig:single-column}
\end{figure*}

\section{Related Work}

\subsection{Sequential Recommendation}
Sequential recommendation models the evolution of user preferences from interaction sequences and has been extensively studied in recommender systems. Early methods such as GRU4Rec~\cite{hidasi2015session}, NARM~\cite{li2017neural}, and STAMP~\cite{liu2018stamp} rely on recurrent or attention-based architectures to capture sequential dependencies, while later models including Caser~\cite{tang2018personalized}, NextItNet~\cite{yuan2019simple}, SASRec~\cite{kang2018self}, TiSASRec~\cite{li2020time}, and BERT4Rec~\cite{sun2019bert4rec} further improve the modeling of local patterns and long-range dependencies through convolution or self-attention. More recently, BSARec~\cite{shin2024attentive} strengthens sequence representation learning with improved architectural design. Although these methods achieve strong performance, most of them still model heterogeneous temporal patterns within a single representation space, making it difficult to explicitly disentangle preference signals at different temporal scales. In contrast, WPGRec emphasizes multi-resolution modeling to represent such temporal patterns more explicitly.

\subsection{Graph-based Recommendation}
Graph-based recommendation incorporates collaborative signals by propagating information over user--item interaction graphs or session-level transition graphs~\cite{Feng2019AGCN,ZhuEtAl2024GiffCF,wang2019neural}. Representative methods include PinSage~\cite{ying2018graph}, LightGCN~\cite{he2020lightgcn}, SR-GNN~\cite{wu2019session}, GC-SAN~\cite{xu2019graph}, SGL~\cite{wu2021self}, TGSRec~\cite{fan2021continuous}, and DGRec~\cite{yang2023dgrec}. These studies show that collaborative structure is an important complement to purely sequential signals and can substantially improve representation quality. Nevertheless, most existing graph-based methods mainly focus on structural propagation and aggregation, with limited attention to the differences among sequential dynamics at different temporal scales. By comparison, WPGRec performs graph propagation over temporally aligned subband representations, enabling scale-consistent collaborative enhancement.

\subsection{Frequency and Time--Frequency Modeling}
Frequency-domain and time--frequency modeling has recently attracted growing interest in sequential recommendation~\cite{zhou2022filter,kim2025diff}. FEARec~\cite{du2023frequency} introduces frequency-enhanced modeling to capture periodic preference patterns, while WaveRec~\cite{HeoKim2025WaveRec} explores wavelet transforms for localized multi-resolution sequence representation. These works collectively show that frequency-aware modeling offers a useful complement to conventional time-domain approaches. They also suggest that explicitly separating signals across different frequency bands can provide a more interpretable view of user preference dynamics. These studies suggest that user behavior contains informative patterns distributed across different frequency bands, and they highlight the value of time--frequency analysis for characterizing non-stationary and localized temporal variation. However, existing methods remain largely sequence-centric and make limited use of collaborative graph information, leaving the joint modeling of multi-resolution representations and graph propagation insufficiently explored. Different from prior frequency-aware recommenders, WPGRec unifies stationary wavelet packet decomposition, subband-wise graph propagation, and energy--spectral-flatness-aware fusion within a single framework.

\section{Method}
The overall architecture of WPGRec is illustrated in~\autoref{fig:single-column}. The core idea of WPGRec is to first decompose each behavior sequence into aligned multi-resolution subbands, then inject collaborative information within each subband through graph propagation, and finally adaptively fuse the resulting representations. In this section, we first formalize the recommendation setting, and then describe the proposed framework from five aspects: problem setup, SWPT-based subband construction, subband-level temporal summarization, subband-wise graph propagation, and gated fusion.

\subsection{Problem Setup}
We model recommendation on a user--item bipartite graph
\begin{equation}
G=(U\cup I,E),
\label{eq:graph}
\end{equation}
where $U$ and $I$ denote the user set and item set, respectively, and $E$ is the interaction edge set.
In this graph, each edge represents an observed interaction between a user and an item.
For each user $u\in U$, the observed chronological behavior sequence is
\begin{equation}
x_u=\{i_1,\ldots,i_{T_u}\},\quad i_t\in I,
\label{eq:seq}
\end{equation}
where $T_u$ is the sequence length and $i_t$ is the item interacted with at time step $t$.
This sequence is the basic temporal signal from which the model infers user preference evolution.

Each item $i$ is associated with an embedding vector $\mathbf e_i\in\mathbb R^d$, where $d$ denotes the embedding dimension.
By indexing the item embeddings according to the order of items in $x_u$, we obtain the sequence embedding matrix
\begin{equation}
\mathbf X_u\in\mathbb R^{T_u\times d},\qquad
\mathbf X_u(t,:)=\mathbf e_{i_t}.
\label{eq:xu}
\end{equation}
Here, $\mathbf X_u(t,:)$ denotes the $t$-th row of $\mathbf X_u$, namely the embedding of the item interacted with at time step $t$.
Therefore, $\mathbf X_u$ can be viewed as the continuous vector representation of the user behavior sequence, and serves as the sequential input to the subsequent multi-resolution modeling module.
Given the prefix sequence $\{i_1,\ldots,i_{T_u-1}\}$, the target is to predict the next item $i_{T_u}$.

\subsection{SWPT Subband Representation}
To reduce boundary artifacts introduced by finite-length sequences, we first apply symmetric extension and append $\tau$ learnable boundary tokens on both sides:
\begin{equation}
\mathbf X_u'=[\tau\text{ tokens},\,\mathbf X_u,\,\tau\text{ tokens}],\qquad
T_u'=T_u+2\tau.
\label{eq:boundary}
\end{equation}
Here, $\tau$ controls the number of added boundary tokens at each side, and $T_u'$ is the resulting extended sequence length.
The purpose of this step is to make the later wavelet decomposition more stable near sequence endpoints, where boundary distortion would otherwise be more severe.

We then apply a full-tree undecimated Stationary Wavelet Packet Transform (SWPT) with depth $\ell$ along the temporal axis:
\begin{equation}
\{Z_u^{(b)}\}_{b=1}^{B}
=\mathrm{SWPT}_{\ell}(\mathbf X_u';\psi),\qquad B=2^\ell,
\label{eq:swpt}
\end{equation}
where $\psi$ denotes the wavelet family, $\ell$ is the decomposition depth, and $B=2^\ell$ is the number of resulting subbands.
The index $b\in\{1,\ldots,B\}$ denotes the subband identity.
Each subband representation
$Z_u^{(b)}\in\mathbb R^{T_u'\times d}$
has the same temporal length as the extended sequence, which means that temporal positions remain aligned across subbands.
This equal-length and shift-invariant property is particularly important in our setting, because it allows later graph propagation and fusion to operate on subband representations without introducing additional scale misalignment.

For each subband $b$, we further compute its energy:
\begin{equation}
E_u^{(b)}=\frac{1}{T_u'}\sum_{t=1}^{T_u'}\|Z_u^{(b)}(t,:)\|_2^2,
\label{eq:energy}
\end{equation}
where $E_u^{(b)}$ measures the average signal magnitude of the $b$-th subband for user $u$.
In addition, we compute the spectral flatness $\mathrm{SFM}_u^{(b)}$ along the temporal axis.
While energy reflects how strong a subband is, spectral flatness characterizes whether its spectrum is more structured or more noise-like.
These statistics are not used as final features by themselves; instead, they serve as descriptors that will later guide adaptive cross-subband fusion.

\subsection{Subband Attention}
After obtaining the subband representations, we summarize the temporal content within each subband into a compact user-level vector.
Let
$z_{u,t}^{(b)}=Z_u^{(b)}(t,:)$
denote the feature vector at time step $t$ in subband $b$ for user $u$.
We then use additive attention to aggregate these token-level features:
\begin{equation}
p_u^{(b)}=\sum_{t=1}^{T_u'} a_{u,t}^{(b)}\,z_{u,t}^{(b)},
\label{eq:attn}
\end{equation}
where $a_{u,t}^{(b)}$ is the normalized attention weight assigned to position $t$ in subband $b$.
Intuitively, $a_{u,t}^{(b)}$ determines how much the corresponding temporal position contributes to the final subband summary.
In this way, the model can assign larger weights to more informative positions, instead of treating every time step equally.

The resulting vector $p_u^{(b)}\in\mathbb R^d$ is the summarized representation of user $u$ in subband $b$.
By stacking these user-level summaries over all users, we obtain
\begin{equation}
\mathbf X_U^{(b)}=[p_u^{(b)}]_{u\in U}\in\mathbb R^{|U|\times d}.
\label{eq:xub}
\end{equation}
Here, $\mathbf X_U^{(b)}$ is the user-side feature matrix for subband $b$, where each row corresponds to one user summary.
This matrix is then used as the user input to the graph propagation stage.

\subsection{Subband Graph Propagation}
For each subband, we construct an initial node feature matrix by concatenating the user-side subband summaries and the item embedding matrix:
\begin{equation}
H^{(b,0)}=
\begin{bmatrix}
\mathbf X_U^{(b)}\\
\mathbf E_I
\end{bmatrix},
\quad
\mathbf E_I=[\mathbf e_i]_{i\in I}\in\mathbb R^{|I|\times d}.
\label{eq:init}
\end{equation}
Here, $\mathbf E_I$ denotes the matrix of all item embeddings, and $H^{(b,0)}$ is the initial node representation for graph propagation in subband $b$.
Its upper block corresponds to users, while its lower block corresponds to items.
In other words, for each subband we build a separate graph feature space in which both user and item nodes are represented.

Based on this initialization, we perform Chebyshev graph filtering for $L_g$ layers:
\begin{equation}
H^{(b,l)}=\sum_{k=0}^{K}T_k(\tilde L)\,H^{(b,l-1)}\Theta_k^{(l)},
\label{eq:cheby}
\end{equation}
where $\tilde L$ is the normalized graph Laplacian, $T_k(\cdot)$ denotes the $k$-th Chebyshev polynomial, and $\Theta_k^{(l)}$ is the learnable filter parameter for order $k$ at layer $l$.
The parameter $K$ controls the polynomial order, i.e., how many spectral components are considered in each propagation step, while $L_g$ controls the number of graph propagation layers.
Through this formulation, collaborative information is propagated over the user--item graph separately within each subband, rather than after different temporal components have already been merged.
This is the key mechanism that makes graph enhancement band-consistent in WPGRec.

After the final propagation layer, we denote the user rows and item rows in $H^{(b,L_g)}$ by $h_u^{(b)}$ and $h_i^{(b)}$, respectively.
These vectors are the graph-enhanced representations of user $u$ and item $i$ in subband $b$.
They will subsequently be fused across subbands to form the final representations used for prediction.

\subsection{Energy--SFM Gated Fusion}
Once subband-specific propagated representations are obtained, the next step is to combine them into final user and item representations.
For the user side, we construct a descriptor
$\phi_u^{(b)}=[E_u^{(b)},\mathrm{SFM}_u^{(b)},\ell]$
for each subband.
Here, $E_u^{(b)}$ is the subband energy, $\mathrm{SFM}_u^{(b)}$ is the spectral flatness, and $\ell$ indicates the decomposition depth used to generate the subbands.
Together, these quantities summarize the strength and spectral characteristic of each subband and provide the input signal for adaptive gating.

Based on this descriptor, the gate weights are computed as
\begin{equation}
g_u^{(b)}=\mathrm{softmax}(\mathrm{MLP}(\phi_u^{(b)})).
\label{eq:gate}
\end{equation}
The scalar $g_u^{(b)}$ represents the relative importance assigned to subband $b$ for user $u$.
Because the gate weights are normalized by the softmax function, the contributions of all subbands are directly comparable and sum to one.

The final user representation is then obtained by weighted fusion over all graph-enhanced subband representations:
\begin{equation}
z_u'=\mathrm{LN}\!\left(\mathbf W_f\sum_{b=1}^{B}g_u^{(b)}h_u^{(b)}\right).
\label{eq:user_fuse}
\end{equation}
Here, $\mathbf W_f\in\mathbb R^{d\times d}$ is a shared projection matrix and $\mathrm{LN}$ denotes layer normalization.
The weighted sum aggregates user information across subbands according to their learned importance, and the projection plus normalization further maps the result into the final representation space.

For the item side, we do not use user-specific gates.
Instead, item representations are fused by simple averaging:
\begin{equation}
z_i'=\mathrm{LN}\!\left(\mathbf W_f\frac{1}{B}\sum_{b=1}^{B}h_i^{(b)}\right).
\label{eq:item_fuse}
\end{equation}
This yields the final item representation $z_i'$ by combining graph-enhanced item features across all subbands.
As a result, WPGRec produces final user and item representations, $z_u'$ and $z_i'$, that jointly encode multi-resolution temporal patterns and band-consistent collaborative information.

\begin{algorithm}[t]
\caption{Workflow of WPGRec}
\label{alg:wpgrec}
\footnotesize
\begin{algorithmic}[1]
\Require Graph $G=(U\cup I,E)$, item embeddings $\{\mathbf e_i\}$, wavelet family $\psi$, depth $\ell$ ($B=2^\ell$), boundary length $\tau$, Chebyshev parameters $(K,L_g)$
\For{each epoch}
  \ForAll{$u\in U$}
    \State Build $\mathbf X_u$ from $x_u$ and obtain $\mathbf X_u'$ by boundary handling (Eq.~\eqref{eq:xu}, Eq.~\eqref{eq:boundary})
    \State Perform SWPT on $\mathbf X_u'$ to obtain $\{Z_u^{(b)}\}_{b=1}^{B}$ (Eq.~\eqref{eq:swpt})
    \For{$b=1$ \textbf{to} $B$}
      \State Compute $E_u^{(b)}$, $\mathrm{SFM}_u^{(b)}$, and the attention summary $p_u^{(b)}$ (Eq.~\eqref{eq:energy}, Eq.~\eqref{eq:attn})
    \EndFor
  \EndFor
  \For{$b=1$ \textbf{to} $B$}
    \State Stack $\mathbf X_U^{(b)}=[p_u^{(b)}]_{u\in U}$ and initialize $H^{(b,0)}$ (Eq.~\eqref{eq:xub}, Eq.~\eqref{eq:init})
    \For{$l=1$ \textbf{to} $L_g$}
      \State Update $H^{(b,l)}$ by Chebyshev graph filtering (Eq.~\eqref{eq:cheby})
    \EndFor
    \State Read out $h_u^{(b)}$ and $h_i^{(b)}$ from $H^{(b,L_g)}$
  \EndFor
  \State Compute gate weights $\{g_u^{(b)}\}_{b=1}^{B}$ for each user (Eq.~\eqref{eq:gate})
  \State Fuse subband-specific user representations to obtain $z_u'$ (Eq.~\eqref{eq:user_fuse})
  \State Fuse subband-specific item representations to obtain $z_i'$ (Eq.~\eqref{eq:item_fuse})
  \State Compute $s(u,i)$ and optimize $\mathcal L$ (Eq.~\eqref{eq:score}, Eq.~\eqref{eq:loss})
\EndFor
\end{algorithmic}
\end{algorithm}

\subsection{Training Objective}
Given the final user and item representations, we define their compatibility score as
\begin{equation}
s(u,i)=\langle z_u',z_i'\rangle.
\label{eq:score}
\end{equation}
Here, $\langle z_u',z_i'\rangle$ denotes the inner product between the fused user representation and the fused item representation.
A larger score indicates that item $i$ is more compatible with user $u$ under the learned representation space.

Based on this score, the full-softmax probability over the whole item set is defined as
\begin{equation}
p(i\mid u)=\frac{\exp(s(u,i))}{\sum_{j\in I}\exp(s(u,j))}.
\label{eq:prob}
\end{equation}
This formulation normalizes the score of the target item against all candidate items in the item set, which is consistent with the full-ranking evaluation protocol adopted in our experiments.

For a mini-batch $\mathcal B$, the training objective is
\begin{equation}
\mathcal L=
-\sum_{(u,i_u^+)\in\mathcal B}\log p(i_u^+\mid u)
+\lambda_1\mathcal L_{\mathrm{gate}}
+\lambda_2\|\theta\|_2^2,
\label{eq:loss}
\end{equation}
where $i_u^+$ denotes the ground-truth next item for user $u$, $\theta$ denotes all trainable parameters, $\lambda_1$ controls the gating regularization term $\mathcal L_{\mathrm{gate}}$, and $\lambda_2$ is the coefficient for $\ell_2$ regularization.
The first term is the full-softmax cross-entropy loss for next-item prediction, the second term regularizes the gating mechanism, and the third term controls model complexity.
The full training workflow is summarized in Algorithm~\autoref{alg:wpgrec}.

\subsection{Complexity Analysis}
Let $N=|U|+|I|$, $B=2^\ell$, $T$ be the average sequence length, and $d$ be the embedding size.
Under this notation, the dominant per-epoch costs come from four parts.
The first part is the user-side SWPT and attention computation, which costs $O(|U|BTd)$.
The second part is subband-wise Chebyshev graph propagation, which costs $O(BL_gK|E|d)$.
The third part is the projection operation, which costs $O(Nd^2)$.
The fourth part is the full-softmax computation over the item set, which costs $O(|\mathcal{B}||I|d)$.
Therefore, the total time complexity is
\begin{equation}
\mathrm{Time}=O\!\left(|U|BTd+BL_gK|E|d+Nd^2+|\mathcal{B}||I|d\right),
\label{eq:complex_time}
\end{equation}
where each term corresponds to one of the four stages above.

The overall space complexity is
\begin{equation}
\mathrm{Space}=O\!\left(L_g(K+1)d^2+|E|+L_gNBd\right).
\label{eq:complex_space}
\end{equation}
Here, the first term mainly comes from the learnable Chebyshev filter parameters, the second term stores the graph structure, and the third term is dominated by the intermediate node representations across graph layers and subbands.
This complexity analysis shows that WPGRec introduces extra multi-resolution computation through subband decomposition and propagation, while keeping the overall complexity in a form that scales predictably with the graph size, sequence length, and decomposition depth.

\begin{table*}[!t]
\caption{Detailed performance of 7 baselines and WPGRec on 4 datasets. The best results are in \textbf{boldface}, and the second-best are \underline{underlined}. ``Improv.'' reports the relative improvement over the best baseline under the full-ranking protocol.}
\centering
\resizebox{\linewidth}{!}{
\begin{tabular}{l c c c c c c c c c c}
\toprule
Dataset & Metric
& BERT4Rec & SGL & WaveRec & FEARec & BSARec & TGSRec & DGRec & WPGRec & Improv. \\
\midrule
\multirow{4}{*}{Beauty}
& HR@10
& 0.0728 & 0.0851 & 0.0894 & 0.0966 & 0.0983 & 0.1079 & \underline{0.1085} & \textbf{0.1197} & +10.3\% \\
& NDCG@10
& 0.0387 & 0.0359 & 0.0552 & 0.0581 & 0.0595 & 0.0672 & \underline{0.0687} & \textbf{0.0742} & +8.0\% \\
& HR@20
& 0.1051 & 0.0998 & 0.1271 & 0.1335 & 0.1337 & 0.1481 & \underline{0.1486} & \textbf{0.1583} & +6.5\% \\
& NDCG@20
& 0.0465 & 0.0471 & 0.0679 & 0.0677 & 0.0686 & 0.0745 & \underline{0.0771} & \textbf{0.0813} & +5.4\% \\
\midrule
\multirow{4}{*}{Sports}
& HR@10
& 0.0412 & 0.0432 & 0.0529 & 0.0584 & 0.0588 & 0.0634 & \underline{0.0643} & \textbf{0.0704} & +9.5\% \\
& NDCG@10
& 0.0225 & 0.0217 & 0.0362 & 0.0331 & 0.0335 & 0.0354 & \underline{0.0369} & \textbf{0.0401} & +8.6\% \\
& HR@20
& 0.0638 & 0.0607 & 0.0884 & 0.0822 & 0.0835 & 0.0879 & \underline{0.0895} & \textbf{0.0991} & +10.7\% \\
& NDCG@20
& 0.0287 & 0.0312 & 0.0405 & 0.0387 & 0.0395 & \underline{0.0421} & 0.0420 & \textbf{0.0439} & +4.2\% \\
\midrule
\multirow{4}{*}{LastFM}
& HR@10
& 0.0496 & 0.0402 & 0.0524 & 0.0582 & 0.0699 & 0.0783 & \underline{0.0785} & \textbf{0.0847} & +7.8\% \\
& NDCG@10
& 0.0261 & 0.0262 & 0.0307 & 0.0321 & 0.0417 & 0.0494 & \underline{0.0495} & \textbf{0.0532} & +7.5\% \\
& HR@20
& 0.0801 & 0.0799 & 0.0808 & 0.0854 & 0.0976 & 0.1035 & \underline{0.1037} & \textbf{0.1197} & +15.4\% \\
& NDCG@20
& 0.0337 & 0.0335 & 0.0349 & 0.0397 & 0.0508 & 0.0572 & \underline{0.0587} & \textbf{0.0642} & +9.4\% \\
\midrule
\multirow{4}{*}{ML-1M}
& HR@10
& 0.2466 & 0.2289 & 0.2527 & 0.2638 & 0.2784 & 0.3424 & \underline{0.3455} & \textbf{0.3542} & +2.5\% \\
& NDCG@10
& 0.1281 & 0.1197 & 0.1392 & 0.1435 & 0.1557 & 0.1968 & \underline{0.1976} & \textbf{0.2004} & +1.4\% \\
& HR@20
& 0.3616 & 0.3591 & 0.3687 & 0.3796 & 0.3860 & \underline{0.4361} & 0.4359 & \textbf{0.4397} & +0.8\% \\
& NDCG@20
& 0.1566 & 0.1498 & 0.1698 & 0.1819 & 0.1828 & 0.2250 & \underline{0.2251} & \textbf{0.2265} & +0.6\% \\
\bottomrule
\end{tabular}}
\label{tab:results_improved}
\end{table*}

\begin{table}[t]
\centering
\caption{Detailed dataset statistics.}
\label{tab:dataset_stats}
\setlength{\tabcolsep}{5.5pt}
\renewcommand{\arraystretch}{1.05}
\begin{tabular}{lcccc}
\toprule
Dataset & \#users & \#items & \#sparsity & \#avg.length \\
\midrule
Beauty & 22,363 & 12,101 & 99.93\% & 8.9 \\
Sports & 25,598 & 18,357 & 99.95\% & 8.3 \\
LastFM & 1,090  & 3,646  & 98.68\% & 48.2 \\
ML-1M  & 6,041  & 3,417  & 95.16\% & 165.5 \\
\bottomrule
\end{tabular}
\end{table}

\section{Experiments}
\label{sec:experiments}

\subsection{Experimental Setup}
\label{sec:subsection}

\noindent\textbf{Datasets and Preprocessing.}
We evaluate WPGRec on MovieLens-1M~\cite{harper2015movielens}, Amazon-Beauty and Amazon-Sports~\cite{ni2019amazon}, and LastFM (HetRec 2011)~\cite{cantador2011hetrec}.
Data preprocessing follows~\cite{shin2024attentive}, including chronological sequence construction, leave-one-out chronological splitting for training, validation, and test sets, and sequence truncation or padding to a fixed maximum length.
The user--item graph is built from training interactions only.
Dataset statistics after preprocessing are reported in ~\autoref{tab:dataset_stats}.

\noindent\textbf{Evaluation Protocol.} We use a full-ranking Top-$K$ protocol, where the ground-truth test item is ranked against the entire item set (excluding items seen in the user’s training history). We report HR@10, HR@20, NDCG@10, and NDCG@20. No sampled negatives are used in either training or evaluation: training uses full-softmax cross-entropy over the entire item set, and evaluation uses full-ranking over all candidate items.

\noindent\textbf{Baselines.}
We compare WPGRec against methods from two paradi-gms: (1) sequence models, including BERT4Rec~\cite{sun2019bert4rec}, BSARec~\cite{shin2024attentive}, FEARec~\cite{du2023frequency}, and WaveRec~\cite{HeoKim2025WaveRec}; (2) graph-based models, including SGL~\cite{wu2021self}, DGRec~\cite{yang2023dgrec}, and TGSRec~\cite{fan2021continuous}. For fair comparison, we re-implement all baselines under the same data splits, train all methods with mini-batch full-softmax cross-entropy, and evaluate them using the same full-ranking Top-$K$ protocol; hyperparameters are re-tuned on the validation set for each method.

\noindent\textbf{Implementation and Training.}
We use $d{=}64$, AdamW (lr $=0.005$, weight decay $=10^{-4}$), batch size $128$, and early stopping on validation NDCG@10 (patience $=10$). 
Training uses full-softmax cross-entropy (no negative sampling) with gating regularization $\lambda_1{=}10^{-5}$. 
Chebyshev propagation uses $K{=}2$ and $L_g{=}2$. 
Grid search is performed on validation sets with $\psi\in\{\text{Coiflets},\text{Symlets}\}$, $\ell\in\{1,2,3\}$, and $\tau\in\{0,2,4,6\}$. 
Results are averaged over three seeds $\{42,43,44\}$ under full-ranking evaluation.

\subsection{Overall Performance}

\noindent\textbf{Comparison with Baselines.}~
As shown in ~\autoref{tab:results_improved}, WPGRec consistently improves over strong baselines across all four datasets.
On \textit{Beauty}, it surpasses the best baseline (DGRec) by $10.3\%$/$6.5\%$ on HR@10/HR@20 and $8.0\%$/$5.4\%$ on NDCG@10/NDCG@20.
On \textit{Sports} and \textit{LastFM}, WPGRec delivers clear gains, with the largest improvements on \textit{LastFM} (e.g., +$15.4\%$ HR@20 and +$9.4\%$ NDCG@20), while on the denser \textit{ML-1M} it maintains a small yet stable lead across metrics.

\noindent\textbf{Result Analysis.}~
Table~\ref{tab:results_improved} shows that WPGRec consistently improves over the strongest baselines across all four datasets and all reported metrics, indicating that its gains are both stable and generalizable. The improvements are particularly notable on the sparse datasets \textit{Beauty}, \textit{Sports}, and \textit{LastFM}: on \textit{Beauty}, WPGRec improves HR@10 from 0.1085 to 0.1197 and NDCG@10 from 0.0687 to 0.0742; on \textit{Sports}, it raises HR@20 from 0.0895 to 0.0991; and on \textit{LastFM}, it achieves the largest gain, improving HR@20 from 0.1037 to 0.1197 and NDCG@20 from 0.0587 to 0.0642, corresponding to relative gains of $15.4\%$ and $9.4\%$, respectively. By contrast, the gains on the denser \textit{ML-1M} dataset are smaller but still consistent, such as the improvement from 0.3455 to 0.3542 on HR@10 and from 0.1976 to 0.2004 on NDCG@10. This pattern suggests that WPGRec is especially effective when user behavior is sparse and temporally heterogeneous, where aligned subband decomposition helps separate signals at different scales, subband-wise graph propagation injects collaborative information without scale mismatch, and the energy--spectral-flatness-aware fusion further emphasizes informative subbands while suppressing noisy ones.

\begin{figure*}[t]
  \centering
  \begin{minipage}[b]{0.4\textwidth}
    \centering
    \includegraphics[width=0.82\linewidth]{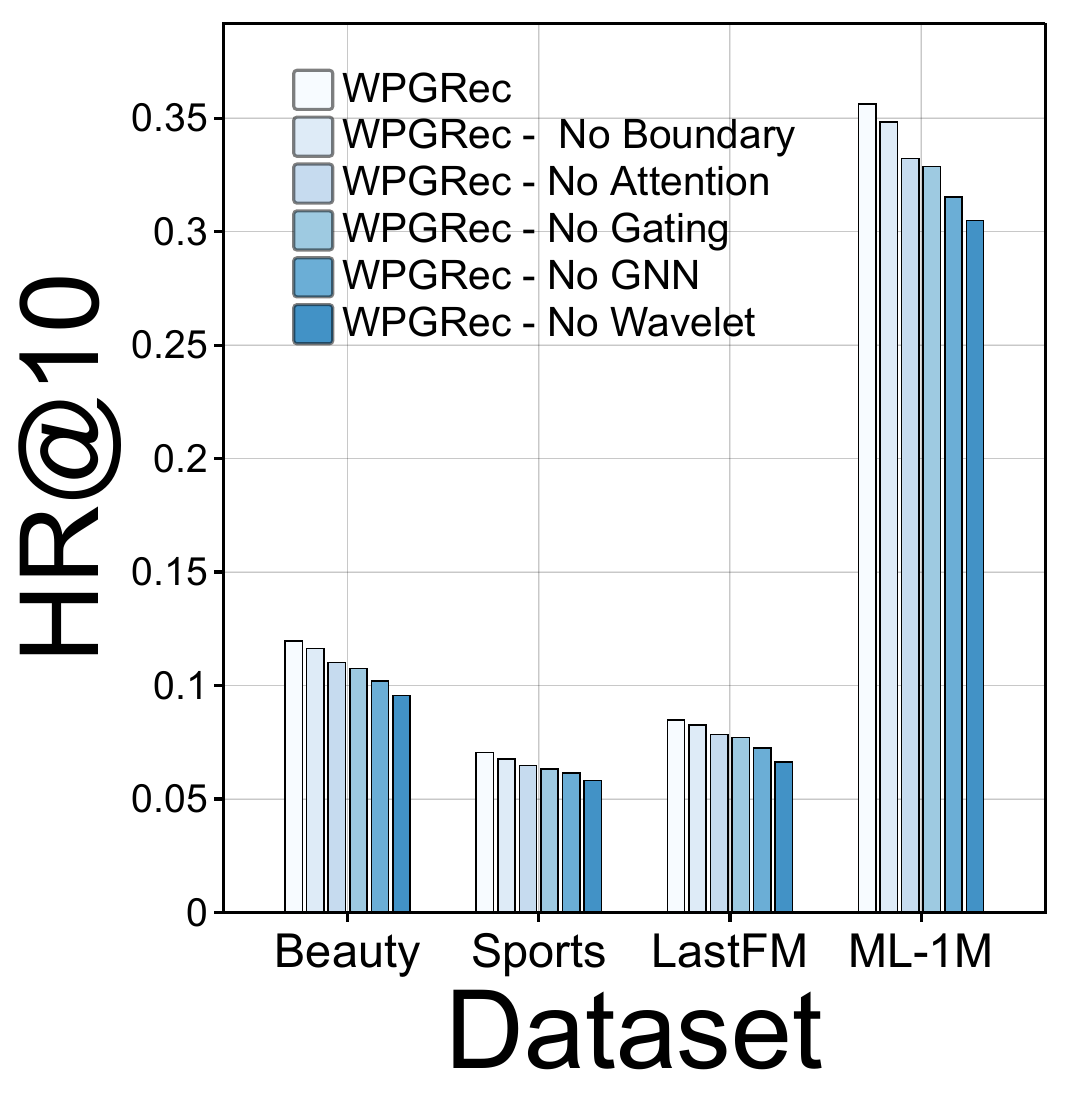}
  \end{minipage}
  \hspace{1.2cm}
  \begin{minipage}[b]{0.4\textwidth}
    \centering
    \includegraphics[width=0.82\linewidth]{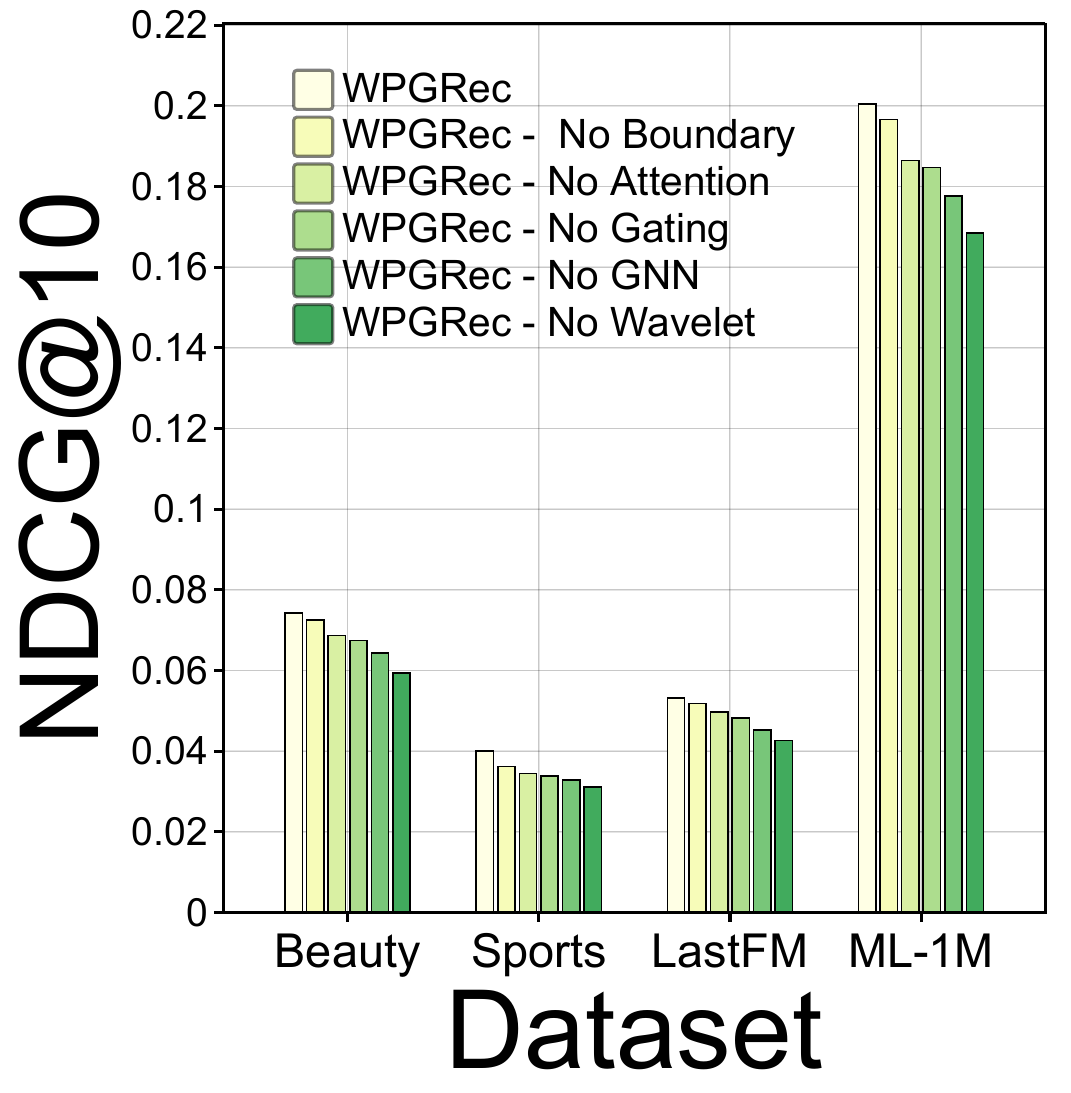}
  \end{minipage}
  \caption{Ablation results (left: HR@10; right: NDCG@10).}
  \label{fig:res}
\end{figure*}

\subsection{Ablation Analysis}
\label{subsec:ablation}

\noindent\textbf{Ablation Setup.}~
To assess the contribution of each component in WPGRec, we evaluate six variants and report HR@10 and NDCG@10 on the four datasets:
(1) the full model WPGRec;
(2) \textbf{No Wavelet}: replace the SWPT module with an identity mapping and set the number of subbands to $B{=}1$, such that the model operates purely in the time domain;
(3) \textbf{No GNN}: remove the subband-wise graph propagation module;
(4) \textbf{No Gating}: replace the cross-subband gating mechanism with uniform averaging;
(5) \textbf{No Attention}: disable the intra-subband attention mechanism; and
(6) \textbf{No Boundary}: remove symmetric extension and boundary tokens by setting $\tau{=}0$.
All variants follow the same training and evaluation protocol as the full model.
Except for the specified module, the remaining architecture and optimization settings are kept unchanged to ensure fair comparison.

\noindent\textbf{Ablation Results.}~
The ablation results in~\autoref{fig:res} show that removing any major component consistently degrades performance, confirming that the gains of WPGRec do not come from a single isolated design choice. Among all variants, removing wavelet-based subband decomposition causes the most noticeable drop, indicating that explicit multi-resolution separation is the foundation of the proposed framework. Disabling subband-wise graph propagation also leads to a clear decline, which suggests that collaborative information is most effective when injected within aligned subband representations rather than after temporal patterns have already been mixed. Replacing the adaptive gating mechanism or removing intra-subband attention results in moderate but consistent deterioration, showing that both adaptive fusion and temporal summarization contribute to the final ranking quality. In addition, removing boundary handling causes smaller yet still stable drops, implying that reducing boundary distortion helps maintain more reliable subband representations, especially for short or sparse sequences. Overall, these results demonstrate that multi-resolution decomposition, band-consistent graph propagation, adaptive fusion, and boundary-aware representation learning jointly contribute to the robustness of WPGRec.

\subsection{Sensitivity Analysis}
\label{subsec:sensitivity}

\noindent\textbf{Sensitivity Setup.}~
To examine the sensitivity of WPGRec to the gating regularization term, we vary $\lambda_1$ from $10^{-7}$ to $10^{-3}$ on two representative datasets, \textit{Beauty} and \textit{LastFM}. This experiment aims to evaluate whether the effectiveness of WPGRec depends on careful tuning of the gating strength, and to identify a stable operating range of $\lambda_1$ under the full-ranking setting.

\noindent\textbf{Sensitivity Results.}~
Table~\ref{tab:sens_lambda1} shows that WPGRec remains strong over a relatively broad range of $\lambda_1$, with the best overall performance achieved at $\lambda_1{=}10^{-5}$. On \textit{Beauty}, both HR@10 and NDCG@10 increase from 0.1192 and 0.0739 at $10^{-7}$ to 0.1197 and 0.0742 at $10^{-5}$, and remain very close at $10^{-6}$ and $10^{-4}$, suggesting a stable plateau over $10^{-6}\!\sim\!10^{-4}$. A similar pattern is observed on \textit{LastFM}, where HR@10 and NDCG@10 rise from 0.0840 and 0.0527 to 0.0847 and 0.0532, and then slightly decline to 0.0836 and 0.0521 at $10^{-3}$. These results suggest that overly small $\lambda_1$ provides insufficient regularization for the gating mechanism, whereas excessively large $\lambda_1$ may over-concentrate gate weights and slightly hurt performance. Overall, the performance gains of WPGRec are stable and do not depend on delicate hyperparameter tuning.

\begin{table}[t]

\centering
\caption{Sensitivity analysis of the gating regularization coefficient $\lambda_1$ (mean over 3 seeds).}

\label{tab:sens_lambda1}
\small
\setlength{\tabcolsep}{3.8pt}
\renewcommand{\arraystretch}{1.05}
\resizebox{\columnwidth}{!}{%
\begin{tabular}{lcccc}
\toprule
$\lambda_1$ & Beauty HR@10 & Beauty NDCG@10 & LastFM HR@10 & LastFM NDCG@10 \\
\midrule
$10^{-7}$ & 0.1192 & 0.0739 & 0.0840 & 0.0527 \\
$10^{-6}$ & 0.1195 & 0.0741 & 0.0844 & 0.0530 \\
$10^{-5}$ & \textbf{0.1197} & \textbf{0.0742} & \textbf{0.0847} & \textbf{0.0532} \\
$10^{-4}$ & 0.1194 & 0.0740 & 0.0843 & 0.0529 \\
$10^{-3}$ & 0.1187 & 0.0734 & 0.0836 & 0.0521 \\
\bottomrule
\end{tabular}%
}

\end{table}

\section{Conclusion}
We presented WPGRec, a wavelet packet guided framework for sequential recommendation that unifies scale-aligned subband modeling with band-consistent collaborative structure injection. By integrating stationary wavelet packet decomposition, subband-wise Chebyshev graph propagation, and energy--spectral-flatness-aware fusion, WPGRec is able to preserve heterogeneous temporal patterns while incorporating high-order collaborative signals in a scale-consistent manner. Extensive experiments on four public benchmarks show that WPGRec consistently outperforms strong sequential and graph-based baselines, with particularly notable gains on sparse and behaviorally complex datasets. These results demonstrate the effectiveness of aligning multi-resolution temporal modeling with graph propagation, and highlight the value of subband-level collaborative enhancement for sequential recommendation. Although the current framework has not been optimized for efficiency under larger-scale or online recommendation settings, it provides a promising foundation for future research. In the future, we plan to explore more efficient subband modeling and propagation strategies, as well as adaptive decomposition mechanisms for more scalable and flexible sequential recommendation.

\begin{acks}
This work was primarily supported by the Jilin Provincial Department of Science and Technology (Grant No.~20260102309JC), and in part by the Excellent Young Scientists Fund (Overseas) of the National Natural Science Foundation of China.
\end{acks}

\bibliographystyle{ACM-Reference-Format}

\bibliography{sample-base}
\end{document}